# Unsupervised Learning of Individual Kohn-Sham States: Interpretable Representations and Consequences for Downstream Predictions of Many-Body Effects

Bowen Hou[1], Jinyuan Wu[1], and Diana Y. Qiu[1*]

[1]Department of Mechanical Engineering and Material Sciences,
Yale University, New Haven, CT 06511, USA

**Abstract**

Representation learning for the electronic structure problem is a major challenge of machine learning in computational condensed matter and materials physics. Within quantum mechanical first principles approaches, Kohn-Sham density functional theory (DFT) is the preeminent tool for understanding electronic structure, and the high-dimensional wavefunctions calculated in this approach serve as the building block for downstream calculations of correlated many-body excitations and related physical observables. Here, we use variational autoencoders (VAE) for the unsupervised learning of high-dimensional DFT wavefunctions and show that these wavefunctions lie in a low-dimensional manifold within the latent space. Our model autonomously determines the optimal representation of the electronic structure, avoiding limitations due to manual feature engineering and selection in prior work. To demonstrate the utility of the latent space representation of the DFT wavefunction, we use it for the supervised training of neural networks (NN) for downstream prediction of the quasiparticle bandstructures within the GW formalism, which includes many-electron correlations beyond DFT. The GW prediction achieves a low error of 0.11 eV for a combined test set of metals and semiconductors drawn from the Computational 2D Materials Database (C2DB), suggesting that latent space representation captures key physical information from the original data. Finally, we explore the interpretability of the VAE representation and show that the successful representation learning and downstream prediction by our model is derived from the smoothness of the VAE latent space, which also enables the generation of wavefunctions on arbitrary points in latent space. Our work provides a novel and general machine-learning framework for investigating electronic structure and many-body physics.

[*] diana.qiu@yale.edu

Recently, machine learning (ML) has emerged as a powerful tool in condensed matter and materials physics, achieving substantial progress across areas including the identification of phase transitions[1-7], quantum state reconstruction[8-10], prediction of topological order[11-15], symmetry[16-22] and the study of electronic structure [23-26]. Among these ML applications, exploring the electronic structure of real materials is of particular interest, since it allows for the extension of computationally-intensive predictive quantum theories to understand the physics of larger and more complex systems, such as moire systems [16,27-35] and defect states[36-42], among others. Within atomistic first principles theories, density functional theory (DFT)[43] is the most commonly used approach for studying the electronic structure of materials. In principle, DFT gives accurate descriptions of the ground state charge density, but quantitative prediction of excited-state properties (including bandstructure and other spectroscopic properties) requires the introduction of the concept of elementary excitations, such as quasiparticles from the many-body ground state [44,45]. Nonetheless, DFT can be used as a starting point for many-body calculations, where the wavefunctions within Kohn-Sham DFT are used to construct correlation functions for the excited states[46,47]. Therefore, harnessing the rich information embedded in the Kohn-Sham (KS) wavefunction[43] becomes crucial for downstream ML applications. In this pursuit, a key challenge lies in distilling a succinct representation of the electronic structure while preserving the essential information[48].

In contrast to the notable achievements of ML descriptors for crystal geometry and chemical composition [49-53], the electronic structure of materials remains extremely challenging to learn for several reasons. Firstly, the high dimensionality of the KS wavefunction of real materials creates a complex data structure, making direct pattern detection challenging. Secondly, electronic structures are highly nonlinear and senstive to both the crystal configuration and intricate non-local correlations, making it difficult to develop a general ML model applicable across a broad spectrum of materials. Thus far, ML models for electronic structure have been mostly confined to the study of specific subsets of materials, such as molecules,[54] perovskites,[55-57] or layered transition metal dichalcogenides (TMDs) [58]. In most of these approaches, the ML has focused on prediciton of single valued properties, such as band gaps, largely due to the challenges of identifying well-defined, interpretable, and efficient representations of the electronic structure [59-61]. Recently, the development of operator representations, which capture more nuanced data about the underlying quantum states, has provided new opportunities for the prediciton of full band structures. These techniques involve the physically-informed selection of specific operators as "fingerprints." Examples include the early successful use of energy decomposed operator matrix elements (ENDOME), combined with radially decomposed projected density of states (RAE-PDOS), to predict quasiparticle (QP) band structures[62], and the use of spectral operator representations to predict material transparency[63]. However, the choice of descriptors in such approaches is informed by human physical intuition on the domain science side, inevitably introducing bias. This raises the question of whether more fundamental or generalizable descriptors can be learned, independent of human selection .

Variational autoencoders (VAEs) [64,65], a class of probablistic models that combine variational inference with autoencoders used to compress and decompress high dimensional data, stand out as a promising tool for learning electronic structure in a way that allows for unsupervised training and thus avoids arbitrary feature engineering. The effectiveness of VAE compression has been demonstrated for various applications across condensed matter physics, including quantum state compression[66], detection of critical features in phase transitions[67] and decoupled subspaces[68,69]. In this work, we showcase for the first time that a well-crafted VAE is capable of representing KS-DFT wavefunctions on a manifold within a significantly compressed latent space, which is $10^3 - 10^4$ times smaller than the original input. Importantly, these succinct representations still retain the full physical information inherent in the initial data. To validate the efficacy of the VAE latent space in practical applications, we then build a supervised deep learning model for downstream prediction of k-resolved quasiparticle energies within the many-body GW approximation [70,71], whose only input is the VAE representation of KS states. This model yields a mean absolute error (MAE) of 0.11 eV, which is comparable to the intrinsic numerical error of the GW approach [35,47,72]. Moreover, the VAE inherently predicts a smooth, physically realistic band structure, in contrast with previous models, which required an additional manually-imposed smoothing function to remove unphysical variations[62]. Lastly, we address the interpretability of the VAE representation and show that the success of electronic structure representation learning and downstream GW prediction is derived from the smoothness of the VAE latent space, which corresponds to the smoothness of the wavefunction in k-space. Because of this property, the VAE can serve as a promising wavefunction generator capable of predicting realistic wavefunctions at arbitrary points in latent space, which has the potentital to be used for both wavefunction interpolation in k-space and materials design and discovery across the compositional phase space.

**VAE for electronic structure representation**

As Fig. 1 (a) shows, the architecture of our VAE consists of the two complementary sets of NN: the encoder ($e_\theta$) and the decoder ($d_\theta$). The encoder maps the input of high-dimensional KS wavefunction moduli $|\phi_{n\vec{k}}(\vec{r})| \in \mathbb{R}^{R_x \times R_y \times R_z}$ to a low-dimensional vector ($\vec{\mu} \in \mathbb{R}^p, \vec{\sigma} \in \mathbb{R}^p$) of variational mean and deviation in the latent space. Here, $p \ll R_x \times R_y \times R_z$, where $p$ represents the dimension of the latent space, and $R_{x,y,z}$ is the dimension of the real-space wavefunction $\phi_{n\vec{k}}(\vec{r})$ in the unit cell, where the wavefunction is indexed by the band *n* and the crystal momentum (k-point) $\vec{k}$. Conversely, the decoder, mirroring the structure of the encoder, has an opposite function, which is to reconstruct the input wavefunction from its low-dimensional latent vector ($\vec{\mu}, \vec{\sigma}$). In essence, the VAE can be conceptualized as an "information bottleneck" for electronic structure, where the encoder acts as a data refinement process, discarding redundant information and noise from the input wavefunction. Meanwhile, the decoder ensures that vital physical information is still preserved for wavefunction recovery. Then, since the real space KS state is essentially a spatial distribution with local pattern, convolutional neural networks (CNN) [73-76] provide a practical framework for the first two layers of the encoder and decoder. Additionally, by imposing a distribution of latent variables close to a standard normal distribution $N(0, I)$ during training, the VAE is capable of a smooth mapping from the connected KS states $|n, \vec{k}\rangle$ and $|n, \vec{k} + \Delta\vec{k}\rangle$ to nearby points in the latent space, enabling generative power after training. Here, we define the total VAE loss function as follows:

$$\mathcal{L} = \frac{1}{n}\sum_{n\vec{k}}^{T} |||\phi_{n\vec{k}}| - d_\theta(e_\theta(|\phi_{n\vec{k}}|))||^2 + \beta \cdot \frac{1}{n}\sum_{n\vec{k}}^{T}\sum_{j=1}^{p}\frac{1}{2}(-1 + \sigma_{n\vec{k}}^{(j)^2} + \mu_{n\vec{k}}^{(j)^2} - \log\sigma_{n\vec{k}}^{(j)^2})$$

(1)

, where the first term is the mean squared error (MSE) of the reconstructed wavefunctions, and the second term is the Kullback-Leibler (KL) divergence, which forces the latent space to approach a $N(0, I)$ distribution. The parameter $\beta$ is the weight of KL divergence, tuning the degree of regularization of the latent space. $T$ is the total number of KS states in the training set and, and $j$ is an index in the latent space. By feeding DFT wavefunctions to a well-trained VAE model, we can interpret the vector $(\vec{\sigma}, \vec{\mu})$ in latent space as a low-dimensional effective representation associated with the individual KS state. The details of the VAE model are given in the SI.

**Downstream machine learning for GW band structures:**

Due to its ground-state nature, DFT calculations often yield inaccurate single-particle band structures and tend to underestimate the band gap of semiconductors compared to experiment[77-79]. To incorporate the many-electron correlation effects that are missing in DFT and obtain accurate quasiparticle (QP) bandstructures, one can replace the effective single-particle exchange correlation potential in DFT with a non-local and energy-dependent self energy calculated within the GW approximation in many-body perturbation theory [46,70,80,81]. This approach proves highly effective for computing high-accuracy QP bandstructures in a wide range of materials, spanning from weakly to moderately correlated quantum systems[46]. However, in practice, constructing a GW self-energy, even for small systems[72,80,82-85], is much more computationally expensive than DFT, and this remains a bottleneck to the broader adoption of the GW approach for high throughput studies. Therefore, in the context of understanding materials' excited-state properties, a natural approach is to use low-fidelity techniques like density functional theory (DFT) to try and predict the results of high-fidelity, computationally intensive many-body calculations. Currently, most ML work in this field focus on the use of indirect predictors for model training, using, for instance, crystal geometrical structure, chemical composition, and the DFT bandgap as input [59-61]. These methods, however, are generally limited to single value prediction of the GW band gap, and only effective within specific subsets of materials, such as inorganic solids[86]. Operator fingerprint methods like ENDOME and RAD-PDOS extend predictive capability to k-resolved band structures, but the selection of features in these models heavily relies on human intuition, introducing inevitable bias and unphysical wiggles in band structures, which are then smoothed in post processing[62]. There have also been ML models dedicated to predicting the screened dielectric function [87,88], which can speed up GW calculations but don't provide information that is generalizable to different systems.

Due to the challenging nature of capturing non-local frequency-dependent correlations in the electronic structure, we select the downstream prediction of GW bandstructures as a proof of principle of the effectiveness of our VAE latent vector representation of the KS wavefunctions. We develop a supervised deep NN on top of the VAE latent space to predict many-body GW corrections. The goal of this NN is to successfully learn the diagonal part GW self-energy $\Sigma_{n\vec{k}}^{GW}$ [70,71]:

$$\Sigma_{n\vec{k}}^{GW} = \frac{i}{2\pi\Omega}\sum_{\vec{G}\vec{G}'}\sum_{\vec{q}}^{BZ}\sum_{m}^{all}\int_{-\infty}^{\infty}d\omega' W_{\vec{G}\vec{G}'}(\vec{q},\omega')\frac{\rho_{m\vec{k}-\vec{q}}^{n\vec{k}}(\vec{G})\rho_{m\vec{k}-\vec{q}}^{n\vec{k}*}(\vec{G}')}{\omega+\omega'-\epsilon_{m,\vec{k}-\vec{q}}+i\eta\cdot\text{sgn}(\epsilon_{m,\vec{k}-\vec{q}}-\mu)} \quad (2)$$

where $\vec{G}$ is a reciprocal lattice vector; $\vec{q}$ is the difference between any two k-vectors $\vec{k}-\vec{k}'$ and is integrated over the Brillouin zone (BZ); $\epsilon$ is KS-DFT energy, and $\rho_{m\vec{k}-\vec{q}}^{n\vec{k}}(\vec{G}) = \langle n\vec{k}|e^{i(\vec{q}+\vec{G})\vec{r}}|m\vec{k}-\vec{q}\rangle$. $W_{\vec{G}\vec{G}'}$ is the screened Colomb interaction calculated within the random phase approximation, and $\omega$ is the frequency dependence of self energy. Notably, the calculation of the GW self energy includes a sum over infinite bands, *m*. In practice, the sum over states is treated as a convergence parameter, and the number of bands included in the summation is of the same order as the number of reciprocal lattice vectors $\vec{G}$ included in $W$.

The diagonal self-energy matrix element for a specific state $|n,\vec{k}\rangle$ shown in Eq.(2) can be expressed as: $\Sigma_{n\vec{k}}^{GW} = f(\phi_{n\vec{k}}, \varepsilon_{n\vec{k}}, \vec{\phi}, \vec{\varepsilon}, \rho)$, where $\phi_{n\vec{k}} \in \mathbb{C}^{R_x \times R_y \times R_z}$ and $\vec{\phi} \in \mathbb{C}^{N \times R_x \times R_y \times R_z}$ are the KS state vectors; $\rho \in \mathbb{R}^{R_x \times R_y \times R_z}$ is the charge density; $N$ is the total number of occupied and unoccupied states $|n', \vec{k}'\rangle$ used in GW calculations. $\varepsilon_{n\vec{k}} \in \mathbb{R}^1$ and $\vec{\varepsilon} \in \mathbb{R}^N$ are the corresponding DFT eigenvalues. Here, we note that $\vec{\phi}$ includes $\phi_{n\vec{k}}$ and $\vec{\varepsilon}$ includes $\varepsilon^{nk}$, but we include both terms explicitly in the function $f$ to make later steps more transparent. The output $\Sigma_{n\vec{k}}^{GW} \in \mathbb{R}^1$ is the GW self energy. Due to the closed-form expression of the self-energy, mathematically, we expect that a simple dense NN can acquire an understanding of the non-linear mapping from the KS wavefunction and energies to $\Sigma^{n\vec{k}}$, which consists of the mapping of $(\mathbb{C}^{(N+1)\times(R_x \times R_y \times R_z)}, \mathbb{R}^{R_x \times R_y \times R_z}, \mathbb{R}^{(N+1)} \to \mathbb{R}^1)$[89,90].

However, two significant challenges prevent the application of a simple NN model such as $\hat{\Sigma}_{n\vec{k}}^{GW} = \hat{f}_{NN}(\phi^{n\vec{k}}, \varepsilon^{n\vec{k}}, \vec{\phi}, \vec{\varepsilon}, \rho, \vec{\eta})$, where $\hat{\Sigma}_{n\vec{k}}^{GW} \in \mathbb{R}^1$ is the NN predicted self-energy, and $\vec{\eta}$ are the parameters of the model. (i) Due to the high computational cost of the GW algorithm, only a limited subset of GW energies near the Fermi level can be exactly calculated and used for the supervised learning training set, denoted as $\vec{\Sigma}_{n\vec{k}}^{train} \in \mathbb{R}^{N_{train}}$. As a result, $N_{train} \ll (N+1) \times (R_x \times R_y \times R_z)$, and overfitting is inevitable. (ii) The "curse of dimensionality" makes it formidable to learn the nonlinear mapping from a high-dimensional sparse $\vec{\phi}$ wavefunction space to the self-energy $\vec{\Sigma}_{n\vec{k}}^{train}$ space. To address these challenges, we adopt the manifold assumption[91-94] that the DFT electronic wavefunction in real space can be modeled as lying on a low-dimensional manifold. Additionally, we assume that two DFT wavefunctions mapped to nearby points on the manifold should have comparable contributions to the final GW energy corrections. If these two assumptions hold true, then the VAE is ideal for downstream prediction of GW self energies.

Here, to capture the non-local correlation, the pseudobands approximation[95-97] is further employed for all states. That is, DFT wave functions with close energies are summed into effective super states (see SI). We assume that the manifold assumption also applies to the super states and charge density, and VAE are used to further remove redundant information in them. Eventually, our semi-supervised, physics-informed model reads:

$$\hat{\Sigma}_{n\vec{k}}^{GW} = \hat{f}_{NN}(e_\theta(\phi_{n\vec{k}}(\vec{r})), \varepsilon_{n\vec{k}}, e_{\theta sup}\left(\vec{\phi}_{sup}(\vec{r})\right), \vec{\varepsilon}_{sup}, e_{\theta\rho}(\rho), \vec{\eta}) \quad (3)$$

, where $e_\theta$, $e_{\theta sup}$ and $e_{\theta\rho}$ are encoders exclusively trained for KS states, super states and charge density respectively. The schematic workflow of our model is shown in Fig. 1 (b). The output of the NN is the predicted GW diagonal self energies $\hat{\Sigma}_{n\vec{k}}^{GW} = \hat{\varepsilon}^{GW} - \varepsilon^{DFT}$, and the inputs are the wavefunction of $|n, \vec{k}\rangle$, super bands corresponding to occupied and unoccupied states on a uniform k-grid, and the ground state energies. These inputs are identical to the input of an explicit GW calculation to eliminate bias due to feature selection.

**Results**

To benchmark the predictive power of our model, we select 302 materials from the Computational 2D Materials Database (C2DB)[98-100], which was also used in the work of Knøsgaard et al[62], for training and validation. The dataset includes both metals and semiconductors across all crystal systems (we note that previous ML for GW prediction in this database was restricted to the subset of semiconducting materials) with the number of atoms ranging from 3 to 4. For the unsupervised VAE training, our dataset is comprised of 68,384 DFT electronic states sampled on a 6×6×1 uniform k-grid for 302 2D materials, which are randomly split into 90% training set and 10% test set. We include the same number of conduction states as valence states for each material in the VAE training, ensuring a comprehensive understanding of the electronic structure of both occupied and unoccupied states. The details of GW calculations for supervised training are provided in the SI. The entire GW energy dataset is randomly partitioned into two subsets: 10% (2201 electronic states) is allocated as the test set, while the remaining 90% (19801 electronic states) is designated as the training set. Here, we completely withhold another small subset of 30 materials from the training set for the VAE (see SI for further details), including three monolayer TMD materials MoS$_2$, WS$_2$ and CrS$_2$ used to demonstrate the effectiveness of the model.

Fig. 2(a) shows three DFT electronic wavefunctions, A, B, and C, for monolayer MoS$_2$ within the x-y plane of the unit cell, which correspond to the dark blue circles shown in Fig. 2(e). Fig. 2(b) shows the VAE-reconstructed wavefunction of states A, B and C, which are nearly identical to the original wavefunction after recovery from the low-dimensional latent space with high coefficients of determination $r^2$ of 0.96, 0.94 and 0.91 respectively. Fig. 2(c) shows the VAE variational mean vector $\vec{\mu}$ for states A, B and C respectively. Overall, the $r^2$ of the VAE-reconstructed wavefunction is 0.92 across the test set when compared with the amplitude of the original DFT wavefunction. The results demonstrate that the original electronic wavefunction in the high-dimensional $\mathbb{R}^{(R_x=40)\times(R_y=30)\times(R_z=30)}$ real space can be effectively compressed by 1200 times into a representing vector in $\mathbb{R}^{p=30}$ that still preserves the vital information needed for reconstruction. In addition, our autonomously determined representation is over 100 times more compact than previous electronic fingerprint approaches[62].

Fig. 2(d) shows the comparison between the GW corrections calculated explicitly and predicted by the downstream ML model, where orange (blue) dots represent the test (training) sets. The model yields a MAE of 0.06 eV ($r^2 = 0.96$) and 0.11eV ($r^2 = 0.94$) for the training and test set respectively, confirming that the representation of the DFT wavefunctions learning by the VAE contains sufficient information to describe the non-local GW self-energy. Here, the training set consists of 90% of all data points in the database composed of both metallic and insulating 2D materials with different symmetries. In addition, Fig. 2(e) shows the ML predicted GW band structure of monolayer MoS$_2$, which agrees remarkably with results obtained by explicit GW calculation (red dots in the bandstructure). Due to the generative power of the VAE latent space[64], even in the absence of electronic states along the high-symmetry path and MoS$_2$ in the training set, our model can accurately predict a smooth GW bandstructure along $\Gamma - M - K - \Gamma$ for MoS$_2$. In contract with previous approaches that apply a smoothing function or simple interpolation of the GW band structures [62], our method guarantees continuity in the ML GW band structures through the interpolation of high-dimensional wavefunctions, which is a significantly more complex challenge. Fig. 2(f) shows how each individual input affects the accuracy of the GW NN model. We find that excluding the latent space representation of the wavefunction $|n, \vec{k}\rangle$ in training significantly reduces the $r^2$ value by 0.5, directly showing the importance of VAE representation of individual KS states in GW prediction (see SI for details about comparative benchmark). We note that GW calculations can also be accomplished through the self-consistent Sternheimer equation, utilizing solely the occupied electronic states[101-103], so in principle, inclusion of empty state information through the use of super bands is not strictly necessary. Here, excluding the super band states encoding the empty states reduces the $r^2$ value by 0.05.

**Interpretability and Generative Power**

Next, to understand the generative power of our model, we open up the black box of the VAE and explore the meaning of the latent space obtained from unsupervised learning. We utilize a 3D t-distributed stochastic neighbor embedding (tSNE) to visualize the VAE latent space for the electronic structure. As depicted in Fig. 3(a), the circles linked by green dashed lines represent for the latent points of the first valence band states along the Γ-K-M-Γ high symmetry path of monolayer $MoS_2$. These points form a continous and enclosed trajectory in the latent space, corresponding to the smooth closed path in the k-space shown in the lower bandstructure of Fig. 3(b). For comparison, the latent trajectories from two other TMD monolayers $WS_2$ (blue) and $CrS_2$ (red), are also shown in Fig. 3(a). The similarity in the electronic structures of these three TMDs (see Fig. 3(b)) are encoded in the similarities of the paths in latent space. To further demonstrate the generative power of the VAE, Fig. 3(b) shows the continous evolution of 15 VAE generated real-space KS wavefunction moduli from M-K in the first valence band of monolayer $MoS_2$. The generated wavefunctions are constructed using the VAE decoder, which processes sampled points along a smooth curve from K to M in the latent space, denoted by the blue arrows in Figure 3(a). Notably, even though the VAE training set is only comprised of states sampled on a uniform k-point grid, and these three TMDs are entirely excluded from the training set, the generated wavefunction in Fig.3(b) can still have a high $r^2$ around 0.9, showing the VAE accurately generates wavefunctions that the training has never seen. The underlying reason is that the three trajectories in Fig. 3(a) lie on the smooth latent surface learned by the well-trained VAE, where similar states are mapped to neighboring points in the latent space. Therefore, the smoothness and regularity of the VAE latent space can be used to quantify the extrapolative and generative power of our model and suggests that any sampled points from the learned latent surface are physically meaningful and can be reconstructed back to a wavefunction. Thus, the current VAE model can also serve as an effective generator of electronic wavefunctions.

Additionally, to explain the success of our model in predicting GW corrections at arbitrary k-points, we first investigate how the supervised training at certain k-points affects the GW prediction at untrained k-points in the BZ. Fig.3 (c) illustrate the average GW MAE for different k-points, as predicted by a dense NN that is only trained with GW energies at the Γ point (left panel) or a single finite momentum k point (right panel). Notably, the prediction error tends to be lower for k-points in proximity to the trained k-point, contrasting with higher errors for those farther away from the trained k-point. This trend originates from the smoothness of the VAE latent space, indicating that two states that are close in the latent space should contribute comparably to the GW corrections, as they are also expected to be close in the k-space as shown in Fig.3 (b). This explains the capability of our model to accurately predict k-resolved GW band structures when trained with the limited and uniform k-grid data.

Finally, to further verify the "manifold assumption" for electronic structures in real materials and explore how GW energies correlate to the VAE-coded representation, we apply the 3D t-SNE on top of the latent mean $\vec{\mu}$ space for all 22002 states with GW labels, whose colors are coded to their calculated GW-correction labels, as Fig. 3(d) shows. The t-SNE analysis provides direct evidence for the manifold assumption: the VAE effectively maps the latent $\vec{\mu}$ space of the electronic wavefunction to a smooth spiraling manifold even for different materials. More intriguingly, even though the unsupervised VAE learning procedure is entirely independent of GW labels, the magnitude of the GW corrections exhibit a distinct pattern distributed across the width of the manifold, as opposed to a random distribution. As a result, the t-SNE analysis serves as a crucial validation, demonstrating that unsupervised representation learning can effectively capture the inherent statistical correlation between GW energies and the electronic wavefunction. Therefore, the unsupervised learning VAE plays a role as pre-learning, which can significantly lower the barrier for the subsequent supervised learning of GW self energies.

**Conclusion**

In summary, we demonstrate that a properly designed VAE model can unsupervisedly learn KS DFT wavefunctions, compressing them as a low-dimensional latent space, in a way that preserves fundamental information needed for downstream prediction of excited-state properties. Since our model autonomously determines the crucial information for preservation through the unsupervised reconstruction of wavefunction data, it can establish a low-dimensional representation that avoids limitations due to feature engineering and selection in prior work. Our VAE model achieves a $r^2$ of 0.92 when reconstructing the wavefunctions in the test set. To further test the effectiveness of the VAE representation of KS states, we train a supervised dense NN for downstream prediction of GW self-energies on top of the latent space of KS states. Notably, we feed our NN the latent space of the charge density, DFT wavefunctions and DFT energies, which also correspond to the physical inputs to an explicit GW calculation, avoiding any human bias in selecting inputs. The resulting model demonstrates a remarkably low MAE of 0.11 eV on a test set and can be used to predict both arbitrary k-points and materials held back from the training set. While other ML models have been used to predict GW bandstructure, the main advantage of our approach lies in our ability to interpret the smoothness of the latent space in relation to the completeness of our training set across both k-space and the space of material chemical and structural composition. This eliminates unphyiscal wiggles seen in previous models and allows us to confidently evaluate the generalizability of our model to states outside the training set improving its generative power. The smooth evolution of KS states in k-space can be mapped to a smooth trajectory in the latent space, and the sampled points from this continous trajectory are physically meaningful and can also be reconstructed, enabling the generation of wavefunctions (and related physical observables) at uncalculated points at no additional cost. As a result, despite being trained solely on uniformly sampled k-points, the k-resolved GW band structures can be accurately predicted by the NN. Here, in this first demonstration, we focus on predicting GW bandstructures of 2D materials, but we expect this framework to be generalizable to other crystalline systems and downstream applications, since it does not rely on the selection of specific features.

## ACKNOWLEDGEMENT

We gratefully acknowledge helpful discussions with Mr. Xingzhi Sun at Yale University. This work was primarily supported by the U.S. DOE, Office of Science, Basic Energy Sciences under Early Career Award No. DE-SC0021965. D.Y.Q. acknowledges support by a 2021 Packard Fellowship for Science and Engineering from the David and Lucile Packard Foundation. Excited-state code development was supported by Center for Computational Study of Excited-State Phenomena in Energy Materials (C2SEPEM) at the Lawrence Berkeley National Laboratory, funded by the U.S. DOE, Office of Science, Basic Energy Sciences, MaterialsSciences and Engineering Division, under Contract No. DE-C02-05CH11231. The calculations used resources of the National Energy Research Scientific Computing (NERSC), a DOE Office of Science User Facility operated under contract no. DE-AC02-05CH11231; the Advanced Cyberinfrastructure Coordination Ecosystem: Services & Support (ACCESS), which is supported by National Science Foundation grant number ACI-1548562; and the Texas Advanced Computing Center (TACC) at The University of Texas at Austin.

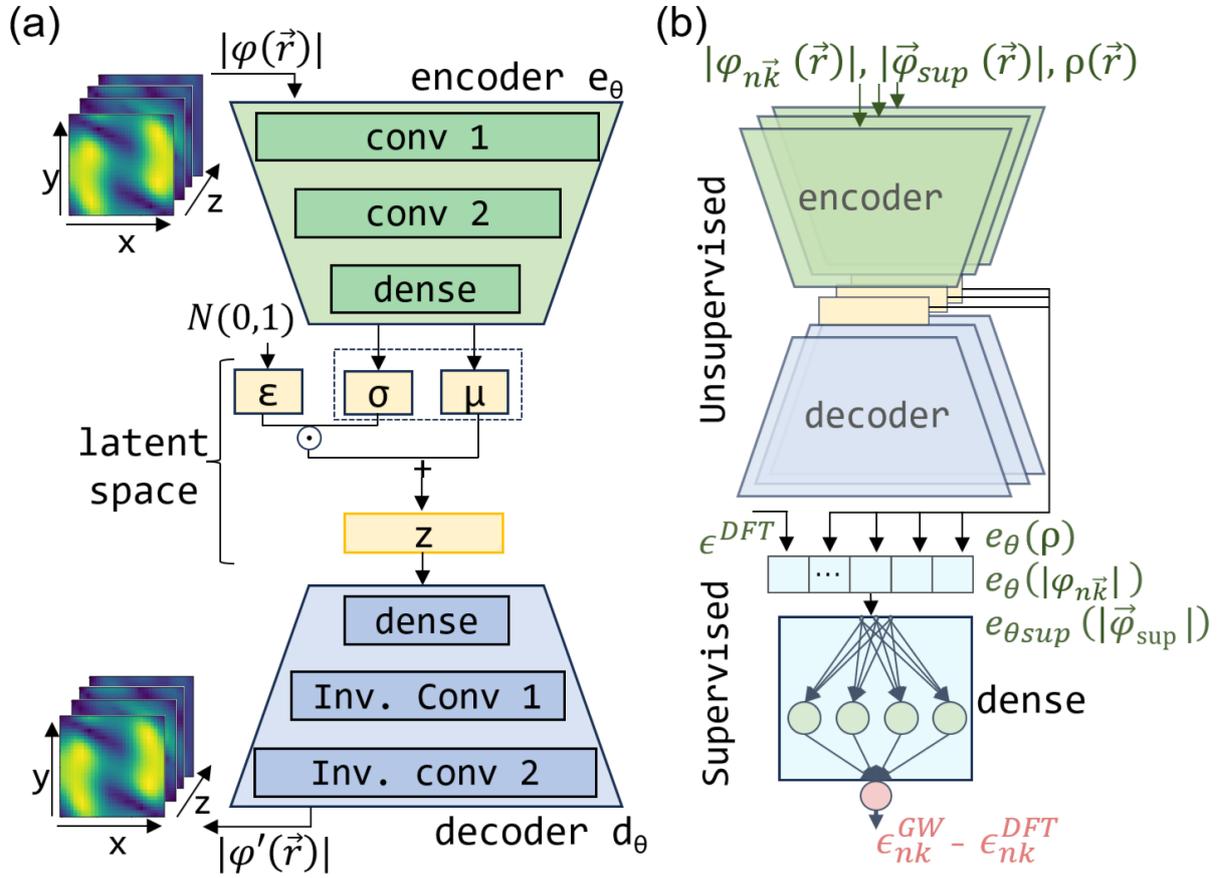

FIG. 1 (a) Schematic of the VAE. The encoder (green trapezoid) consists of two convolutional neural networks (CNN) and one flattened dense layer, mapping a real space wavefunction to a latent space vector of variational mean $\vec{\mu}$ and variance $\vec{\sigma}$. Z is the sampled latent vector, drawn from a variational Gaussian distribution using a "reparameterization trick"[64]. The decoder (blue trapezoid) has symmetric NN structures. The latent space serves as an information bottleneck for the VAE as its dimensions are only 1/1200 of the input and output (represented by the colormaps of the wavefunction in real space). (b) Schematic of the overall semi-supervised learning model, including both the unsupervised VAE and supervised dense neural networks (NN). The VAE inputs are the KS wavefunction modulus $|\varphi_{n\vec{k}}(\vec{r})|$, all super states $|\vec{\varphi}_{sup}(\vec{r})|$ and charge density $\rho(\vec{r})$ in real space. The input layer of the supervised dense NN is comprised of DFT energies, denoted as $\varepsilon^{DFT}$, along with low dimensional effective representations of $\varphi_{n\vec{k}}(\vec{r})$, $\varphi_{sup}(\vec{r})$ and $\rho(\vec{r})$ denoted as $e_\theta(\varphi_{n\vec{k}}(\vec{r}))$,

$e_{\theta sup}(\varphi_{sup}(\vec{r}))$ and $e_{\theta \rho}(\rho(\vec{r}))$ respectively. These representations are encoded within the VAE latent space (yellow square) through an encoder with parameters $\theta, \theta_{sup}$ and $\theta_\rho$, which are unsupervisedly trained for all KS wavefunctions, super states and charge density.

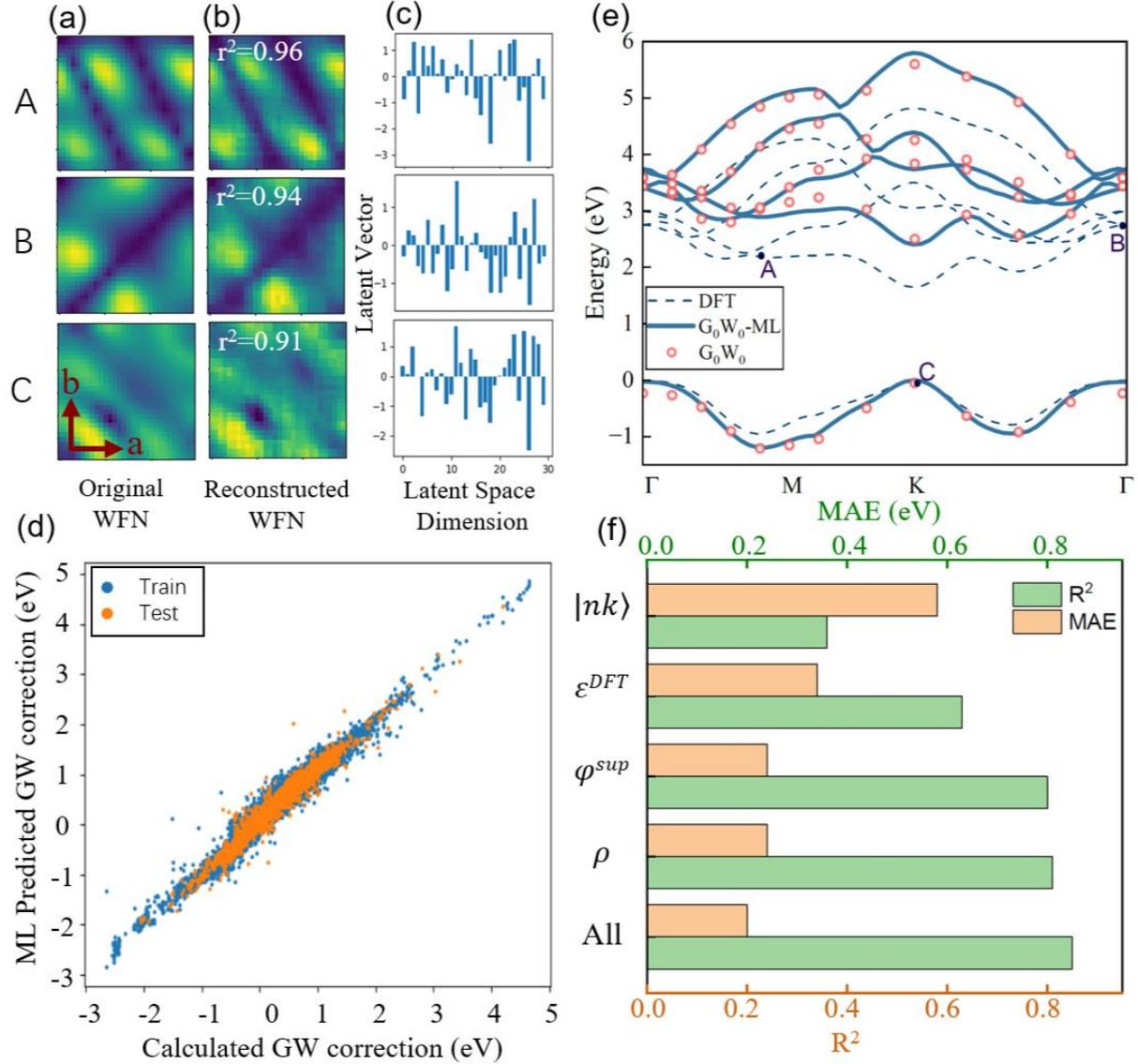

FIG. 2 (a) DFT calculated wavefunction of states A, B and C in MoS$_2$ (see (e)) used as input to VAE encoder. (b) VAE reconstructed wavefunctions of states A, B and C through latent space decoding. (c) Low-dimensional variational mean latent space for states A, B and C. (d) parity plot comparing the exact calculated values (x-axis) to the ML predicted values (y-axis) of the GW correction for individual state. Blue (orange) dots represent training (test) sets. The MAE for the training set and test set are 0.06 and 0.11 eV respectively. (e) ML predicted GW band structures (blue solid curve) and calculated PBE band structures (blue dashed line) for monolayer MoS$_2$. The red circles are the calculated GW energies. (f) Green (yellow) solid bars represent for R$^2$ (MAE) of ML model without utilizing specific information, with the training process spanning 20,000 epochs.

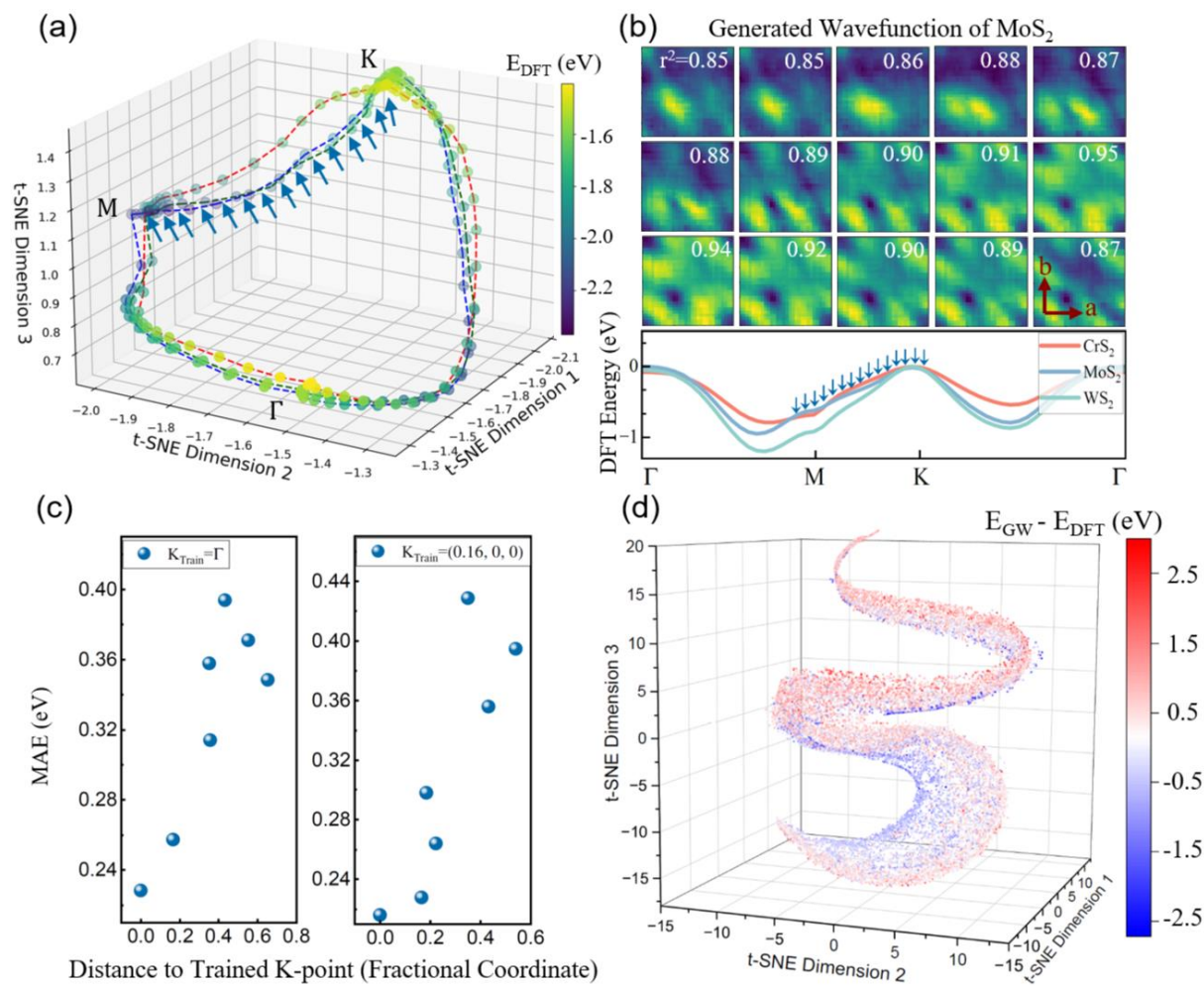

FIG.3 (a) 3D tSNE visulization of the $\vec{\mu}$ latent space for KS wavefunctions in the first valence band along $\Gamma - M - K - \Gamma$ for the TMD monolayers MoS$_2$, WS$_2$ and CrS$_2$, as shown in the lower band structures in (b). The dots (latent points) are mapped from the high-dimensional electronic wavefunction with color coded by DFT energies. The blue, green and red dashed lines connect latent points corresponding to the KS states from the first valence band of MoS$_2$, WS$_2$ and CrS$_2$ respectively. Blue arrows denote latent points of 15 near states along $M - K$ from MoS$_2$, as shown in (b). (b) Generated real-space wavefunction moduli obtained by inputing latent points, which are sampled from the smooth latent curve of monolayer MoS$_2$, into decoder for each KS state from $M - K$ in the first valence band. The inset white number represents the r$^2$ value, showing the high correlation with the calculated wavefunction. The lower figure represents the DFT bandstructres of monolayer MoS$_2$ (Blue), WS$_2$ (Green) and CrS$_2$ (Red). The blue arrows indicate the KS states shown in the upper figures. (c) MAE of GW prediction for different k-points and generative power for k-point interpolation. The model is trained exclusively using GW energies at the $\Gamma$ (left panel) point or $k = (0.1667, 0, 0)$ (right panel). The x-axis is the distance from the trained k-point to the untrained k-points in reciprocal space. (d) 3D t-SNE components of VAE latent $\mu$ vector of 22002 KS wavefunction with GW energy labels, whose color is mapped to the GW correction $\epsilon^{GW} - \epsilon^{DFT}$.